\documentclass[nofootinbib,aps,a4paper,11pt,letterpaper,superscriptaddress,onecolumn,showkeys,times,eqsecnum]{revtex4-2}
\usepackage{amsmath}
\usepackage{amsfonts}
\usepackage{graphicx}
\usepackage{color}
\usepackage{braket}
\usepackage{dcolumn}
\usepackage{bm,url}
\usepackage[linktocpage]{hyperref}
\usepackage{subfigure}
\usepackage{amsfonts}
\usepackage[left=2.1cm,right=2.1cm,top=2cm,bottom=2cm]{geometry}
\usepackage[usenames,dvipsnames,svgnames]{xcolor}  %% colored text
\usepackage{hyperref}   %% Needs to make hyper reference of any references or citations
\definecolor{oxfordblue}{rgb}{0.0, 0.13, 0.28}
\definecolor{burgundy}{rgb}{0.5, 0.0, 0.13}
\definecolor{darkolivegreen}{rgb}{0.33, 0.42, 0.18}
\definecolor{darkblue}{rgb}{0,0,0.5}
\definecolor{richcarmine}{rgb}{0.84, 0.0, 0.25}
\definecolor{darkblue}{rgb}{0,0,0.5}
\definecolor{bluer}{rgb}{0.00,0.50,0.75}{}
\hypersetup{colorlinks=true, citecolor=red, linkcolor=blue,
urlcolor = magenta, filecolor=magenta}

\begin{document}

\newcommand\be{\begin{equation}}
\newcommand\ee{\end{equation}}
\newcommand\bea{\begin{eqnarray}}
\newcommand\eea{\end{eqnarray}}
\newcommand\bseq{\begin{subequations}} %solo con amsmath
\newcommand\eseq{\end{subequations}}
\newcommand\bcas{\begin{cases}}
\newcommand\ecas{\end{cases}}
\newcommand{\p}{\partial}
\newcommand{\f}{\frac}

\title{Could regular primordial black holes be dark matter? }
\author{\textbf{Mohsen Khodadi}}
\email{khodadi@kntu.ac.ir}
\affiliation{School of Physics, Institute for Research in Fundamental Sciences (IPM),	P. O. Box 19395-5531, Tehran, Iran}
\affiliation{Center for Theoretical Physics, Khazar University, 41 Mehseti Str., AZ1096 Baku, Azerbaijan}
\affiliation{School of Physics, Damghan University, Damghan 3671641167, Iran}

\date{\today}
%%%%%%%%%%%%%%%%%%%%%%%%%%%%%%%
\begin{abstract}
The recent proposal proposed by Paul Davies and colleagues [Phys. Rev. D \textbf{111} (2025) no.10, 103512] that regular primordial black holes (RPBHs) form stable, zero-temperature remnants and could thereby constitute dark matter is critically examined. While the introduction of a fundamental length scale indeed regulates the Hawking temperature, preventing its divergence, we show that the evaporation timescale for such RPBHs is infinite. This result holds generically for analytic regular black hole spacetimes under standard adiabatic and quasi-static evolution. Consequently, RPBHs never actually reach a true remnant state within any finite time, but instead persist as slowly evaporating objects with a non-zero luminosity. When the combined emission from a cosmological population of these near-remnants is considered, the resulting radiation is found to violate stringent observational constraints from the cosmic microwave background and extragalactic gamma-ray backgrounds. Therefore, low-mass RPBHs are not viable dark matter candidates.	
	
\vspace{0.5cm}
\textbf {Keywords:} Regular black holes, Primordial black holes, Dark matter, Hawking radiation
\end{abstract}

\maketitle
\section{Introduction}

Dark matter remains one of the most profound challenges in the standard cosmological model. Despite the remarkable precision of modern cosmology in testing the foundations of this framework, many critical questions remain unresolved. While the existence of dark matter—an invisible form of matter that exerts detectable gravitational influence—is well-supported, its fundamental nature continues to elude explanation, with numerous competing theories proposed (see, e.g., \cite{Arbey:2021gdg} for a recent review and \cite{Bertone:2016nfn} for historical context). Among the candidate explanations for dark matter, the primordial black holes (PBHs) emerged in the early 1970s \cite{Carr:1974nx,Chapline:1975ojl} as a compelling possibility (see Refs.  \cite{Green:2020jor,Carr:2020xqk,Villanueva-Domingo:2021spv,Carr:2024nlv} for more details). 
PBHs, in essence, are black holes theorized to have formed in the early universe through the gravitational collapse of overdensities arising from quantum fluctuations. Alternative formation mechanisms, such as phase transitions, also predict their existence. Recently, a new method has been explored that directly connects the tensor power spectrum to PBH abundance to constrain primordial GWs using limits on PBH abundances \cite{Kumar:2025jfi}.
Unlike stellar-origin black holes, PBHs are not formed through stellar collapse, allowing their masses to span an extraordinarily broad range—from the Planck mass up to the mass of the observable universe enclosed by today’s Hubble horizon. However, their potential abundance is strongly constrained by the diverse and significant impacts they would have had on cosmological evolution.
Heavy PBHs-- typically with masses greater than $10^{23}$g-- could manifest through multiple astrophysical signatures, including gravitational lensing, binary disruption, matter accretion, and coalescence events. These phenomena translate into observable constraints from microlensing surveys, stellar population studies, X-ray observations, and gravitational wave detection. Collectively, these limits suggest that PBHs with masses greater than $10^{23}$g cannot make up all of the dark matter \cite{Auffinger:2022khh}, and the presence of lighter PBHs is crucial. Lighter PBHs have potentially richer physics because they are associated with the Hawking radiation process \footnote{This process is unsuitable for heavy PBHs because their evaporation time scales with mass, eventually exceeding the universe's age. For instance, a solar-mass PBH ($\sim 10^{33}$g) has an evaporation timescale exceeding $10^{67}$ years—orders of magnitude longer than the current universe's age ($\sim 10^{10}$ years).}.

Stephen Hawking's seminal work \cite{Hawking:1974rv} revealed that BHs are not truly black, but instead emit quantum radiation near their event horizons. This profound discovery established that BHs gradually lose mass through this radiation process, raising fundamental questions about their ultimate fate. As evaporation proceeds, the BH's mass decreases while its temperature increases exponentially, amplifying the radiation intensity. This semi-classical process culminates in the BH's complete evaporation, predicted to end in a final explosive emission of high-energy particles with no remnant remaining. In other words, as the BH's mass approaches zero, its temperature diverges toward infinity. This singularity in the semi-classical Hawking radiation framework signals the inevitable breakdown of this description, requiring the inclusion of quantum gravitational effects or new physics to properly characterize the final moments of evaporation. Such corrections may fundamentally modify the terminal phase of BH decay.

PBHs within the ''Hawking radiation window" would emit intense, multi-spectral particle radiation. This characteristic emission may provide the only viable method for constraining their abundance. Notably, this critical mass range corresponds to MeV-band emissions – a relatively unexplored frontier in astrophysics that has recently become the focus of major observational initiatives (e.g., AMEGO, ASTROGRAM) \cite{Engel:2022bgx}. However, the important point in this context is that low-mass PBHs (typically with masses lower than $ 10^{15} g$) would have entirely evaporated via Hawking radiation by the present epoch. Consequently, such light PBHs cannot contribute to the observed dark matter density in the universe. The standard evaporation process raises a profound theoretical challenge: the BH information loss paradox \cite{Polchinski:2016hrw} (see also recent paper \cite{Afshordi:2025nlj}). While quantum field theory rigorously requires information preservation, complete Hawking evaporation appears to irreversibly destroy this information. This contradiction with quantum mechanics' unitary evolution principle reveals a fundamental incompleteness in our current understanding of BH thermodynamics.

Recent work by Paul Davies and colleagues \cite{Davies:2024ysj} has revealed a fundamental connection between two central singularities in BH physics: the divergence in Hawking temperature (as mass approaches zero) and the spacetime singularity at the BH's core. Their analysis demonstrates that these infinities are intrinsically related, suggesting a deeper unification between BH thermodynamics and general relativistic geometry.
Under the assumption that physical BHs are non-singular so call regular BHs (RBHs),  Paul Davies and colleagues have demonstrated that a broad class of such objects exhibit finite maximum temperatures during evaporation. This crucial result implies the existence of slowly evaporating, low-mass RPBHs that can persist as stable remnants-- making them viable candidates for cold dark matter \footnote{A recent series of papers \cite{Calza:2024fzo,Calza:2024xdh,Calza:2025mwn} claims a rich phenomenology arising from the link between dark matter and the singularity problem. Also, recently in Ref. \cite{Dialektopoulos:2025mfz} shown that RPBHs can act as cosmic expansion accelerators. }.

As established in \cite{Hayward:2005gi,Frolov:2014jva}, regular BH models offer a theoretically complete and self-consistent description of black hole evaporation through to its final state. This framework rests on two fundamental insights:

1- The absence of a central singularity necessitates both outer and inner horizons that evolve dynamically under Hawking radiation. These horizons progressively converge until reaching an extremal configuration. 

2- The regularization of the BH's core removes the pathological behavior associated with singularities, allowing for a smooth mathematical description of the horizon merger and ultimate disappearance.

This dual mechanism provides a singularity-free resolution to the complete evaporation process. The existence of an even number of horizons in RBHs \cite{Carballo-Rubio:2019fnb,Bonanno:2020fgp} plays a crucial role in their dynamics. Notably, the presence of a non-extremal inner horizon generically leads to mass inflation instability \cite{Poisson:1989zz,Balbinot:1993rf}. However, recent work \cite{Bonanno:2022rvo} has demonstrated that for RBHs, the inner core instability depends not only on mass inflation but also significantly on Hawking radiation effects, representing a distinctive feature of these geometries. For RBHs (non-singular) provide a robust resolution to the singularity problem, the Cauchy horizon must remain stable against generic perturbations \cite{Carballo-Rubio:2018pmi,DiFilippo:2022qkl}. This stability requirement presents a key challenge for any complete regular BH model \cite{Carballo-Rubio:2019nel,Khodadi:2024efq}. Although substantial progress has been made, major challenges persist in identifying and characterizing well-motivated, physically plausible classes of RBHs. The recent review in Ref. \cite{Carballo-Rubio:2025fnc} summarizes these challenges and presents key research directions that show potential.

Davies and colleagues' claim \cite{Davies:2024ysj} regarding known RBH remnants as dark matter candidates requires careful examination, as their analysis does not account for two crucial factors: dynamical stability against perturbations (particularly of inner horizon), and complete evaporation timescales. The comprehensive analysis present in \cite{Carballo-Rubio:2018pmi} demonstrates that existing regular black hole models fail to constitute self-consistent theoretical frameworks. Specifically, these models cannot provide a complete and physically reliable description of BH evaporation throughout the entire process-- from initial Hawking radiation to final disappearance. 

In the following, after briefly reviewing the proposal in \cite{Davies:2024ysj}, we argue that the evaporation timescale of known RBHs is effectively infinite. As a result, these objects do not create remnants within a finite time frame, indicating that the low-mass RPBHs do not contribute to dark matter. In this regard, it is recommended to review some previous studies, e.g., Ref. \cite{Pacheco:2018mvs}.

\section{Reviewing the proposal for regular PBHs as dark matter}
 In this section, we provide a quick review of the recent paper by Paul Davies and colleagues \cite{Davies:2024ysj} in which it is shown that regular BHs can potentially open a new window for PBH dark matter. By skipping on the details in \cite{Davies:2024ysj}, the authors first have considered the following singular-free 2D dilaton gravity metric
\begin{equation}
\mathrm{d}s^2 = -n(r) \mathrm{d}t^2 + n(r)^{-1} \mathrm{d}r^2 \,,
\end{equation}
%\begin{widetext}
\begin{eqnarray}\label{eqg00}
		n(r) =\frac{1}{3}\left(\frac{m}{l}\right)^{\frac{2}{3}}\ln{\frac{r^2-\left(ml^2\right)^{\frac{1}{3}}r+\left(ml^2\right)^{\frac{2}{3}}}{(r + \left(ml^2\right)^{\frac{1}{3}})^2}} +
		\frac{2}{\sqrt{3}}\left(\frac{m}{l}\right)^{\frac{2}{3}}\arctan
		{\left(\frac{2r- (ml^2)^{\frac{1}{3}}}{\sqrt{3}(ml^2)^{\frac{1}{3}}}\right)},
	\end{eqnarray}
%\end{widetext}
and have computed the flux and Hawking temperature correspond to it as follows

\begin{equation}\label{eqflux}
\frac{dm}{dt} =\frac{\kappa^2}{48\pi}= \frac{m^2 r_{0}^2}{48\pi (r_{0}^3 + ml^2)^2} \,,
\end{equation}
and
\begin{equation}\label{T}
T_{H} = \frac{\kappa}{2 \pi} = \frac{mr_{0}}{2\pi(r_{0}^3 + ml^2)}.
\end{equation} respectively. Here, $m$, $l$ and $r_0$ represent the BH mass, the fundamental length scale, and horizon location respectively. The latter guarantees that an observer falling into the BH  approaches a smooth, constant, maximally curved spacetime, with Ricci scalar, $R_{max} \propto l^{-2}$ \footnote{It is important to note that the boundedness of curvature invariants alone does not guarantee regularity; geodesic completeness must also be considered \cite{Carballo-Rubio:2019fnb}.}. It is worth noting that the aforementioned results are compatible with previous findings obtained via the method of complex paths \cite{Easson:2002tg}. The fundamental length scale $l$ in the Hawking temperature mentioned above causes, unlike the Schwarzschild Hawking temperature, it to reach a maximum; after that, the BH begins to cool, possibly settling as a remnant mass. The flux, \eqref{eqflux}, gives us the evaporation time as follows
\begin{equation}\label{t}
t = 48 \pi \int \frac{(r_{0}^3 + m l^2)^2}{m^2 r_{0}^2} \, dm \,.
\end{equation}
To solve this integral, knowing the exact solution for the horizon position $r_0$ is essential. Although it may seem out of reach, one can estimate the remnant mass by taking the limit as $r\rightarrow 0$, which corresponds to the late stages of BH evaporation. In this way, by expanding $n(r)$ around $r=0$, along with applying an integration constant from the asymptotically flat requirement ($n \rightarrow 1$ as $r \rightarrow \infty$), and keeping the terms up to second order, the lapse function \eqref{eqg00} takes the following form
\begin{equation}
n(r) = 1 - \frac{4 \pi}{3 \sqrt{3}}\left(\frac{m}{l}\right)^{\frac{2}{3}} + \frac{r^2}{l^2} \,,
\end{equation}
Now by solving $n(r)=0$, obtain the horizon location $r_0$
\begin{equation}\label{r0}
	r_{0} = \frac{l}{3} \sqrt{4 \sqrt{3} \pi \left(\frac{m}{l}\right)^{\frac{2}{3}} - 9}\,,
\end{equation}
Finally, since the remnant BH  mass $m_{*}$ is calculated in the limit $r_{0} \rightarrow 0$, we come to
\begin{equation}
m_* = \frac{9}{8} \sqrt{\frac{\sqrt{3}}{\pi^3}} \,l \,.
	\label{remnantMass}
\end{equation}
To see the role of fundamental length scale parameter $l$ on the remnant BH mass, it is enough to put $r_0=0$ in \eqref{T}, which results in $T_H=0$. While in the absence of $l$, it diverges. In other words, unlike the Schwarzschild BH, the presence of the regularization parameter $l$ in the spacetime prevents the Hawking temperature from growing uncontrollably by reducing mass.

\begin{figure}[h]
	\includegraphics[width=0.48\textwidth]{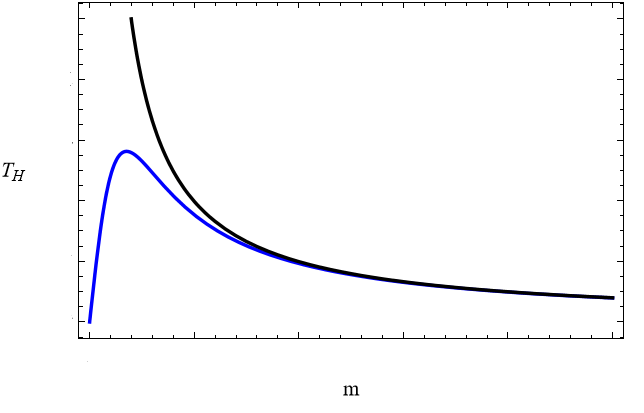}
	\caption{Qualitative behavior of the Hawking temperature correspond to 4D Schwarzschild BH (black curve) and 4D regular BHs (blue curve) in terms of the mass.}
	\label{fig:TH}
\end{figure}

The authors in \cite{Davies:2024ysj} established their proof of concept for an analytical (non-singular) 2D dilaton gravity metric and generalized it to 4D-RBHs, incorporating various forms of mass functions $M(r)$ enriched with a fundamental length scale $l$. Specifically for the Bardeen with mass function $ \frac{mr^3}{(r^2+(2ml^2)^{2/3})^{3/2}}$ \cite{Bardeen}, Hayward with mass function $\frac{mr^3}{r^3+2ml^2}$ \cite{Hayward:2005gi}, Fan-Wang $\frac{mr^3}{(r+l)^3}$ \cite{Fan:2016hvf}, and Dymnikova with mass function $m\big(1-\exp[-\frac{r^3}{2ml^2}]\big)$ \cite{Dymnikova:1992ux}. All in all, the key argument proposed in \cite{Davies:2024ysj} is that since the Hawking temperature of regular spacetimes has a maximum value--unlike Schwarzschild, which is unbounded (Fig. \ref{fig:TH} )-- so BH evaporation in the end leaves a remnant BH mass. It is important in the sense that, unlike the conventional picture of Hawking radiation and evaporation process, it allows the low-mass PBHs to constitute a significant fraction of the dark matter in our universe. However, in the next section, we will discuss and show that the evaporation time of 4D-RBHs, in essence, is infinite, which is physically worthless. 

As a final comment in this section, it is necessary to mention this neglected point from the authors' perspective that 
by inserting $r_0$ from \eqref{r0} into \eqref{t} and its solve, obtains a long expression which in the limit $m\rightarrow m_*$ it diverges. It means that the required time to reach to remnant mass for the 2D dilaton BH at hand is infinite. This is precisely the key point that will be discussed and proven in the next section through an independent analysis for the 4D-RBHs with a general mass function $M(r)$.

\section{Evaporation time of regular BHs}
Generally, the geometry of a spherically symmetric regular BHs can be described by the metric (see e.g., \cite{Dymnikova:1992ux,Bardeen,Fan:2016hvf,Borde:1996df,Hayward:2005gi,Frolov:2017rjz})
\begin{equation}\label{eq:metrictr}
	\text{d}s^2=-e^{-2\phi(r)}F(r)\text{d}t^2+\frac{\text{d}r^2}{F(r)}+r^2(\text{d}\theta^2+\sin^2\theta\, \text{d}\phi^2).
\end{equation}
where $\phi(r)$ and $F(r)$ are two real functions. When convenient, we will alternatively use the notation
\begin{equation}
	F(r)=1-\frac{2 M(r)}{r}.
\end{equation}
$M(r)$, in essence, is the Misner--Sharp--Hernandez quasi-local 
mass~\cite{Misner:1964je,Hernandez:1966zia}. From the Einstein field equations, one immediately finds that the effective energy density corresponding to the metric in Eq. \eqref{eq:metrictr} takes the form $M'(r)/4\pi r^2$, where $M'(r)=\text{d}M(r)/\text{d}r$. This density remains finite at $r = 0$ if and only if $M(r)$ vanishes at least as fast as $r^3$ in the $r \rightarrow 0$ limit. Furthermore, when the dominant energy condition is satisfied, the regularity of the effective energy density guarantees the regularity of the effective pressures \cite{Dymnikova:2001fb}. Without losing generality, for simplicity we suppose that there are only two horizons, outer and inner, whose locations, respectively $r_+$ and $r_-$, are defined by
\begin{equation}
	F(r_\pm)=0.
\end{equation}
By introducing the ingoing Eddington-Finkelstein coordinate $v$,
\begin{equation}
\text{d}v=\text{d}t+\frac{\text{d}r}{e^{-\phi(r)}F(r)},
\end{equation}
the line element~\eqref{eq:metrictr}, re-express as follows
\begin{equation}\label{eq:metric}
	\text{d}s^2=-e^{-2\phi(r)}F(r) \text{d}v^2+2e^{-\phi(r)} \text{d}r \text{d}v+ r^2(\text{d}\theta^2+\sin^2\theta\, \text{d}\phi^2).
\end{equation}
The specialization $M(r)=M$ and $\phi(r)=0$, yields the conventional Schwarzschild metric. Remarkably, the metric in \eqref{eq:metric} admits a representation in terms of the outgoing Eddington-Finkelstein coordinate $u$, where the differential relation $\text{d}u=\text{d}t-\text{d}r/e^{-\phi(r)}F(r)$. 
The line element of \eqref{eq:metric}, openly show that the ingoing radial null curve is determined respectively by the equations
\begin{equation}
\text{d}v=0,\label{eq:ingnull}
\end{equation}
and
\begin{equation}
\frac{\text{d}r}{\text{d}v}=\frac{e^{-\phi(r)}F(r)}{2}.\label{eq:outnull}
\end{equation}
A similar equations can be written for the outgoing radial null curve, too.
The Taylor expansion of Eq.~\eqref{eq:outnull} around $r=r_\pm$ up to the leading order, is given by
\begin{align}
	{\text{d}r\over \text{d}v} &=F(r_\pm)+\frac{e^{-\phi(r_\pm)}}{2}F'(r_\pm)\, (r-r_\pm)\nonumber\\
	&=	\kappa_\pm(r-r_\pm),
\end{align}
where $F(r_\pm)=0$, and the surface gravities at the outer and inner horizons are given by
\begin{equation}
	\label{eq:kappa:pm}
	\kappa_\pm =\frac{e^{-\phi\left(r_\pm\right)}}{2} \left.\frac{\partial F(r,M)}{\partial r}\right|_{r=r_\pm}=\left.-\frac{e^{-\phi(r)}}{2}\left(2M(r)\over r\right)'\right|_{r=r_\pm}.
\end{equation}
The presence of the two horizons, in principle, offers a complete and self-consistent picture of BH evaporation (e.g., see \cite{Frolov:2014jva}). When Hawking radiation is included, the outer and inner horizons gradually approach one another until they merge, at which point the BH becomes extremal. Subsequently, the horizons vanish, leaving behind a nonsingular remnant with finite nonzero mass. However, as we shall demonstrate, this scenario is generally oversimplified, and a more rigorous analysis is necessary.

Typically, the evaporation time has not been systematically investigated, but it is generally assumed to be finite. However, Ref. \cite{Alesci:2011wn} as an exception has been demonstrated that the evaporation time of a loop BH is infinite, ruling out the formation of a remnant. To address the problem, we adopt the standard perspective, which is based on two conservative assumptions: adiabatic and quasi-static conditions. The former, means that the sole relevant dynamical process during evaporation is Hawking radiation, which remains thermal at every stage, with a temperature determined by $T=\frac{\kappa_+}{2\pi}$. The latter tells that the evaporation proceeds as a quasi-static process, with the BH evolving continuously through a sequence of equilibrium states.

According to our hypothesis, the mass loss rate is determined by the Stefan--Boltzmann law,
\begin{equation}\label{eq:Boltz}
	\frac{\text{d}M(v)}{\text{d}v}=-\sigma_{\rm SB}\,T^4(v)A_+^2(v)=-C\kappa_+^4(v)r_+^2(v),
\end{equation}
where $A_+(v)$, $\sigma_{\rm SB}$, and $C$ are the area of the outer horizon, , the Stefan--Boltzman constant and a positive constant, respectively.  Note that
$M$, $\kappa_+$ and $r_+$ have been promoted to dynamical functions of the evaporation time $v$. As a result, the metric \eqref{eq:metric} takes the following form 
\begin{equation}\label{eq:metricv}
	\text{d}s^2=-e^{-2\phi(r,M(v))}F(r,M(v)) \text{d}v^2+2e^{-\phi(r,M(v))} \text{d}r \text{d}v+ r^2(\text{d}\theta^2+\sin^2\theta\, \text{d}\phi^2).
\end{equation}
where the notations $\phi(r,M)$ and $F(r,M)$ denote this fact that these functions, in essence, are mass dependent $M$, so that through $M(v)$ 
these functions have implicitly dependency on the time $v$.
To solve the Eq. ~\eqref{eq:Boltz}, we must take an integral of it, which requires knowing $r_+$ and $\kappa_+$ as functions of $M$. The former, obtain
of $F(r_+,M)=0$. Besides, at extremality limit i.e., $r_+=r_*$ and $M=M_*$, the following condition must satisfy
\begin{equation}\label{eq:extr}
	\left.\frac{\partial F(r,M_*)}{\partial r}\right|_{r=r_\star}=0
\end{equation}
Equation \eqref{eq:kappa:pm} clearly indicates that the surface gravity $\kappa_*$ of an extremal black hole becomes zero unless $\phi(r_*,M_*)$ exhibits a divergence. Consequently, the evaporation rate given in \eqref{eq:Boltz} approaches zero as the black hole reaches the end of its evaporation process. Such an asymptotically vanishing evaporation rate suggests that the total evaporation time could, in fact, be infinite—a conclusion already demonstrated explicitly for near-extremal Reissner–Nordström black holes in \cite{Fabbri:2000es}.

We are now in a position to compute the evaporation time for regular black holes described by the general metric form given in \eqref{eq:metricv}. By invoking the two key assumptions of the standard evaporation picture - namely, adiabatic and quasi-static evolution - we find the black hole's mass \( M \) approaches arbitrarily close to its extremal value \( M_\star \), reaching \( M = M_* + \Delta M \).
Our approach involves integrating Eq.~\eqref{eq:Boltz} from \( M_* + \Delta M \) to \( M_* \) and demonstrating that the corresponding time interval diverges.  
Consider a configuration where the outer horizon radius differs from $r_*$ by an arbitrarily small amount $\Delta r$, with a corresponding infinitesimal mass deviation $\Delta M$ from the extremal value $M_*$, i.e., 
\begin{equation}
r_+=r_*+\Delta r=r_\star\left(1+\epsilon\right),\qquad 0<\epsilon\ll 1
\end{equation}
and 
\begin{equation}
\label{eq:mtoext}
M=M_*+\Delta M=M_*\left(1+\beta\epsilon^\sigma \right)+\mathcal{O}(\epsilon^{\sigma})
\end{equation}
Here, $\Delta M$ parametrized with two real constants $\beta$ and $\sigma>0$.
Due to relation between  $M$ and $r_+$ via $F(r_+,M)=0$, thereby, these constants cannot be arbitrarily values. By expanding $F(r_+,M)=0$ around $\Delta r$ and $\Delta M$ as follows
%%%
\begin{align}
	\label{eq:fexp}
	0=&F(r_*,M_*)+\left.\frac{\partial F}{\partial\,r}\right|_{r_*,M_\star}\Delta r+\left.\frac{\partial F}{\partial\,M}\right|_{r_*,M_\star}\Delta M+\nonumber\\
	&+\left.\frac{1}{2}\frac{\partial^2 F}{\partial r^2}\right|_{r_*,M_*}\Delta r^2+\left.\frac{\partial^2 F}{\partial r\partial M}\right|_{r_*,M_*}\Delta r\,\Delta M+\nonumber\\
	&\left.+\frac{1}{2}\frac{\partial^2 F}{\partial M^2}\right|_{r_*,M_*}\Delta M^2+\dots
\end{align}
one can find their relation. The first and second terms of Eq.~\eqref{eq:fexp} vanishes due to $F(r_+,M)=0$--evaluated at $r_\star$ and $M_\star$-- and Eq.~\eqref{eq:extr}, respectively. As a result, if $\partial F/\partial M\rvert_{r_\star,M_\star}\neq0$, then $\Delta M$ is at least quadratic in $\epsilon$. More generally, $\Delta M$ is of order $\epsilon^n$, where $n$ is the first natural number for which
\begin{equation}
\label{eq:n}
	\left.\frac{\partial^n F}{\partial r^n}\right|_{r_\star,M_\star}\neq0,
\end{equation}
so that $\sigma=n$ and
\begin{equation}
	\beta=-\frac{1}{n!}\left(\left.\frac{\partial F}{\partial M}\right|_{r_*,M_*}\right)^{-1}\left.\frac{\partial^n F}{\partial r^n}\right|_{r_*,M_*}.
\end{equation}
Given that we assume that  $\phi(r,M)$ is finite, thereby, Eq.~\eqref{eq:extr} indicates that $\kappa_*=0$ (see Eq.~\eqref{eq:kappa:pm}). Now, we can parametriz the deviation of $\kappa_+$ from $\kappa_*=0$ with two real constants $\alpha$ and $\gamma>0$.
\begin{equation}
	\label{eq:ktoext}
	\kappa_+=\alpha\epsilon^\gamma+\mathcal{O}(\epsilon^{\gamma}),
\end{equation}
Inserting Eqs.~\eqref{eq:mtoext} and~\eqref{eq:ktoext} in the Stefan-Boltzmann law~\eqref{eq:Boltz}, and the integrate of it, then the evaporation time $\Delta v$ takes the following form
\begin{equation}
	\Delta v=-\frac{M_*}{r_*^2\,C}\frac{\beta\sigma}{\alpha^4}\int_{\epsilon_0}^{0} d\epsilon \,\epsilon^{\sigma-4\gamma-1}.
\end{equation}
As it is clear, $\Delta v$ is finite just provided that 
\begin{equation}\label{eq:fin_time}
\sigma-4\gamma>0.
\end{equation}
On the other hand, by serving the Taylor expansion of $\left.\partial F/\partial r\right|_{r_+}$ in Eq.~\eqref{eq:kappa:pm}, we have
\begin{align}
\label{eq:kexp}
\kappa_+&=\frac{e^{-\phi(r_+)}}{2}\left.\frac{\partial F}{\partial r}\right|_{r_+}\nonumber\\
&=\frac{e^{-\phi(r_+)}}{2}\sum_{i,j=0}^\infty\left.\frac{1}{i!j!}\frac{\partial^{(i+j+1)}F}{\partial^{(i+1)}r\partial^{(j)}M}\right|_{r_*,M_*}r_*^i\epsilon^i\Delta M^j,
\end{align}
where by keeping the leading term in the expansion, it reads as
\begin{equation}
\label{eq:kexp:2}
\kappa_+=\frac{e^{-\phi(r_\star)}}{2}\left.\frac{1}{n!}\frac{\partial^{n}F}{\partial r^n}\right|_{r_\star,M_\star}r_*^{n-1}\epsilon^{n-1}.
\end{equation}
Here $n$ is the same as in Eq.~\eqref{eq:n}. By comparing Eq.~\eqref{eq:kexp:2} and \eqref{eq:ktoext}, it follows that $\gamma=n-1$ and then $\sigma-4\gamma=4-3n$. However, condition $n\geq2$, does not allow the Eq.~\eqref{eq:fin_time} to hold, meaning that $\Delta v$ is infinite. 

By relaxing the assumption $\partial F/\partial M\rvert_{r_*,M_*}\neq0$, Eqs.~\eqref{eq:fexp},~\eqref{eq:fin_time} and~\eqref{eq:kexp:2} give us let to consider of the evaporation time under 
the following three separate cases
\begin{description}
	\item[a) $\bm{\sigma=1}$] Eq.~\eqref{eq:fin_time} becomes $1-4\gamma>0$ i.e., $\gamma<-1/4$. However, the Eq.~\eqref{eq:kexp:2} tells us that, $\gamma$ should be a positive integer which is a disagreement.
	\item[b) $\bm{\sigma<1}$] Eq.~ \eqref{eq:kexp:2} leads to two distinct subcases. First, if $\gamma$ is a positive integer, then Eq. \eqref{eq:fin_time} is immediately violated. Second, if $\gamma$ can be expressed as $\gamma = I + J\sigma$, where $I\geq0$ and $J > 0$ are integers, then we necessarily have $\gamma\geq \sigma$. This implies the inequality $\sigma-4\gamma<-3\sigma<0$, which directly contradicts inequality \eqref{eq:fin_time}.
	\item[ c) $\bm{\sigma>1}$] The term of order $\Delta r^n\sim\epsilon^n$ in Eq.~\eqref{eq:fexp} (with $n$ defined by Eq. \eqref{eq:n} can only be canceled by terms of the form $\Delta r^I\Delta M^J\sim\epsilon^{I+J\sigma}$, where $I,J \geq 1$ are integers satisfying $n=I+J\sigma$. This cancellation condition immediately requires $\sigma\leq n$. Furthermore, the same reasoning leading to Eq.~ \eqref{eq:kexp:2} establishes that $\gamma=n-1$. Consequently, we obtain the inequality $\sigma-4\gamma\leq4-3n<0$, where the final inequality holds because $n\geq2$. This result demonstrates that inequality \eqref{eq:fin_time} cannot be satisfied.
\end{description}
It is evident that all well known regular geometries presented in \cite{Davies:2024ysj} do not satisfy inequality \eqref{eq:fin_time} since they belong to the special case where $\sigma=2$ with $\gamma=1$, corresponding to the third case in the aforementioned classification.

\section{Physical Viability and Observational Constraints}

Our analysis reveals that the formation of a zero-temperature remnant from a RPBH requires an infinite time under the adiabatic and quasi-static assumptions. However, this alone does not rule out the possibility of low-mass RPBHs as dark matter; instead, to evaluate their viability as dark matter candidates, we must consider their behavior within the finite age of the universe, \(t_U \approx 13.8 \, \text{Gyr} \approx 4.35 \times 10^{17} \, \text{s}\).

Let us consider a population of low-mass RBHs with an initial mass that would have led them to the near-extremal regime today. We adopt the Hayward model \cite{Hayward:2005gi} as a representative example, where the mass function is \(M(r) = m r^3 / (r^3 + 2ml^2)\). The extreme mass and horizon radius are \(m_* = r_*/2 = l/\sqrt[3]{4}\). The Hawking temperature and mass loss rate near extremality scale as \cite{Alesci:2011wn,Fabbri:2000es}
\begin{equation}
T_H \propto (m - m_*)^{1/2}, \quad \text{and thus} \quad |\dot{m}| \propto T_H^4 A_+ \propto (m - m_*)^2.
\end{equation}
This scaling, where the evaporation rate vanishes quadratically as \(m \to m_*\), is generic for the well-known models cited in \cite{Davies:2024ysj}. At the first step, let us provide a timescale estimation.
The time \(\Delta v\) to evaporate from a mass \(m = m_* + \Delta m\) down to \(m_*\) is given by integrating \(dm/dv \propto -(\Delta m)^2\). This yields:
\begin{equation}
\Delta v \propto \int_{m_*+\Delta m}^{m_*} \frac{dm}{(\Delta m)^2} \propto \frac{1}{\Delta m}.
\end{equation}
A perfect remnant i.e., $\Delta m=0$ is not possible within a finite timescale.
However, let us take the mass deviation extremely small i.e., near to remnant but not a perfect remnant. In what follows, we show that even a unimaginably small deviation from perfect remnant status leads to physical consequences (continuous evaporation and integrated energy release) that are incompatible with our observations of the universe, thus ruling out these objects as dark matter. In other words, the minuscule difference is physically decisive.

Let's perform the detailed integration to prove that a black hole that is closer to extremality today (smaller final $\Delta m$) has, in fact, emitted more total energy over its lifetime. Before it, note that the problem is not the total energy lost, but how that energy was emitted over time.  We start with the evaporation law for a near-extremal Hayward-type black hole
\begin{equation} \label{m}
\frac{dm}{dv} = -K (m - m_*)^2 
\end{equation}
where: \( K \) is a positive constant,  \( m_* \) is the extremal mass, \( m(v) \) is the mass at time \( v \), and the mass deviation is \( \delta(v) = m(v) - m_* \). We want to calculate the total energy \( E_{\text{total}} \) emitted from an initial time \( v_i \) to the present day \( v_f = v_i + t_U \). This is simply the total mass lost times \( c^2 ~~\text{(square of the speed of light)}\)
\begin{equation}\label{to}
E_{\text{total}} = \left( m(v_i) - m(v_f) \right) c^2 = \left( \delta_i - \delta_f \right) c^2 
\end{equation}
where \( \delta_i = \delta(v_i) \) and \( \delta_f = \delta(v_f) \).
First, we solve the differential Eq. (\ref{m}). Rewrite it in terms of the deviation \( \delta \)
\begin{equation}
\frac{d\delta}{dv} = -K \delta^2,~~~\Longrightarrow \int_{\delta_i}^{\delta} \frac{d\delta'}{(\delta')^2} = -K \int_{v_i}^{v} dv'
\end{equation}
This is a separable equation and leads to the following mass deviation at any time \( v \)
\begin{equation}
\delta(v) = \frac{1}{K (v - v_i) + \frac{1}{\delta_i}} \
\end{equation}
We are interested in the case where the evaporation time is the age of the universe: \( v - v_i = t_U \). At the final time, \( \delta(v_f) = \delta_f \). So,  
\begin{equation}\label{five}
\delta_i = \frac{1}{\frac{1}{\delta_f} - K t_U} 
\end{equation}
Now we plug Eq. (\ref{five}) back into (\ref{to}) for the total energy:
\begin{equation}\label{six}
	E_{\text{total}} = \frac{K t_U \delta_f^2}{1 - K t_U \delta_f} c^2 
\end{equation}
Let us examine Eq. (\ref{six}). From term in the denominator: \( 1 - K t_U \delta_f \) is clear that for the black hole to have been evaporating for time \( t_U \), we must have \( \delta_i > \delta_f > 0 \). From Eq. (5), this requires \( \frac{1}{\delta_f} - K t_U > 0 \), which means \(0< K t_U \delta_f < 1 \). Now, consider what happens as the final deviation \( \delta_f \) gets smaller. The dependence on \( \delta_f \) is dominated by the denominator. As \( \delta_f \) decreases, the denominator decreases faster than the numerator, causing the total energy \( E_{\text{total}} \) to increase. At the first galance, this may seem a bit paradoxical i.e., a black hole which is closer to being a remnant has actually emitted more total energy. But, in essence this paradoxical position is a direct and inevitable consequence of the infinite time required to reach the perfect remnant state.

The successful predictions of Big Bang Nucleosynthesis (BBN) and the precise blackbody spectrum of the Cosmic Microwave Background (CMB) place severe constraints on energy injection from decaying dark matter after \(t \sim 1-10^3\) seconds and \(t \sim 10^{13}\) seconds, respectively \cite{Auffinger:2022khh}. Furthermore, the observed intensity of the extragalactic gamma-ray and neutrino backgrounds tightly constrains the present-day luminosity of any dark matter component. The Fermi-LAT telescope has measured the diffuse gamma-ray background. Now, we are a right postion to compare of our result with the upper limit extracted from Fermi-LAT telescope \cite{LAT} for any dark matter contribution to this background.

For example if we set \( \delta_f = 10^5 \, \text{and}~ 10^4 \, \text{g} \) (i.e., $10^{-8} \%$, and $10^{-10} \%$ away from the perfect remnant mass, respectively) for mass scale \( m_* = 10^{15} \, \text{g} \), then the values of energy injected within lifetime of universe (\( t_U = 4.35 \times 10^{17} \, \text{s} \)) respectively are $\approx9 \times 10^{25} \text{erg}$, and $\approx10^{30} \text{erg}$. Now, let us encounter the observational limits with these values. For a dark matter mass in the range of $10^{15}-10^{16} \text{g}$, the Fermi-LAT data \cite{LAT} constrains the decay lifetime to be: $\tau>10^{28} \, \text{s}$. This is a robust, order-of-magnitude value consistently found in analyses (e.g., see \cite{Cirelli:2010xx}). By translating lifetime bound to luminosity bound, the maximum allowed luminosity for a single RBH of mass $m_*=10^{15}\, \text{g}$ to comply with Fermi-LAT bounds is $L^{max}\lesssim 10 \text{W}$. This means that to avoid producing a detectable gamma-ray background, each individual RBH must have an average luminosity less than roughly 10 Watts. As a result, the allowed total energy per BH over the age of the universe is $\approx 4 \times 10^{25}\, \text{erg}$, meaning that both are above the Fermi-LAT limit (the former slightly above, the latter catastrophically above). The lower $\delta_f$ (more approach to the extreme state), the worse the upper bound violation becomes. There is no reasonable, fine-tuned choice of $\delta_f$ that can bring the RPBH's integrated historical emission down to a level that satisfies the stringent constraints from the diffuse gamma-ray background.

\section{conclusion}
This work has presented a critical assessment of the proposal that low-mass RPBHs can serve as viable dark matter candidates by forming stable, zero-temperature remnants \cite{Davies:2024ysj}. Our analysis leads to the firm conclusion that this proposal is not tenable within the well-defined framework of semi-classical gravity, for two fundamental and interconnected reasons.
First, on theoretical grounds, we have demonstrated that the evaporation timescale for a generic class of analytic RBH geometries is infinite. The mechanism that imposes a maximum on the Hawking temperature—the introduction of a fundamental length scale—also causes the surface gravity to vanish as the black hole approaches its extremal state. This, in turn, causes the mass loss rate to slow asymptotically, following $\mid dm/d\nu\mid \propto (m- m_*)^2$. The integral of this rate from any finite mass deviation $\bigtriangleup m$ to the extremal mass $ m_*$ diverges. Therefore, the purported stable remnant is a mathematical endpoint that is never reached in any finite time. The BH evolution stalls indefinitely in a near-extremal state. Second, and decisively, on phenomenological grounds, this mathematical result translates into a critical physical problem. These slowly evaporating RPBHs would produce a diffuse background of radiation. Integrated over the age of the universe, this energy injection violates stringent observational constraints from present-day extragalactic gamma-ray and X-ray surveys.

It is crucial to emphasize that our conclusion is conditional on the standard semi-classical picture of adiabatic and quasi-static Hawking evaporation. While non-perturbative quantum-gravitational effects could potentially alter the late-time evolution, such models remain speculative and are not part of the current RBH proposal. Within the established framework of regular black holes, the path to remnant formation is kinematically blocked by an infinite timescale, and the resulting physical behavior is observationally excluded.
This reveals the flaw that the adiabatic approximation breaks down. The assumption that the BH evolves slowly through a series of equilibrium states is invalid because the timescale for the final stage of evaporation is infinite. The system cannot maintain equilibrium over an infinite time scale; quantum non-adiabatic effects must become dominant. Thus, while RBHs remain a fascinating subject for theoretical inquiry into the singularity problem, they do not provide a solution to the dark matter mystery. As a concluding remark, the framework of RBHs becomes significantly more complex when accounting for the instability of their regular cores under perturbations within finite timescales, as demonstrated in \cite{Carballo-Rubio:2018pmi}.

 \begin{acknowledgments}
The author thanks the referee for the insightful comments and constructive criticism, which have significantly improved the rigor and clarity of this manuscript.
\end{acknowledgments}

%%%%%%%%%%%%%%%%%%%%%%%%%%%%%%%%%%%%%%%%%%%%%%%%%%%%%%%%%%%%%

\end{document}